\documentclass[11pt]{article}

\usepackage{cite}
\usepackage{subfig}
\usepackage{helvet}
\usepackage{amsmath}
\usepackage{amssymb}
\usepackage{setspace}
\usepackage{setspace}
\usepackage{graphicx}
\usepackage{empheq}
\usepackage{subfig}
\usepackage{soul}

\def\be{\begin{equation}}
\def\ee{\end{equation}}

\usepackage[a4paper,top=3cm,bottom=3cm,left=2.5cm,right=2.5cm,bindingoffset=0mm]{geometry}

\begin{document}

\titlepage                                                    

 \vspace*{0.5cm}
\begin{center}                                                    
{\Large \bf On the Consistent Use of Scale Variations in PDF \\ \vspace{0.3cm}
Fits and Predictions}\\

\vspace*{1cm}
                                                   
L. A. Harland--Lang$^{1}$, R. S. Thorne$^{2}$, \\                                                 
                                                   
\vspace*{0.5cm}
${}^1$Rudolf Peierls Centre, Beecroft Building, Parks Road, Oxford, OX1 3PU, UK \\                                                    
${}^2$Department of Physics and Astronomy, University College London, WC1E 6BT, UK                                              
                                                    
\vspace*{1cm}                         

\begin{abstract}

\noindent  We present an investigation of the theoretical uncertainties in parton distribution functions (PDFs) due to missing higher--order corrections in the perturbative predictions used in the fit, and their relationship to the uncertainties in subsequent predictions made using the PDFs. We consider in particular the standard approach of factorization and renormalization scale variation, and derive general results for the consistent application of these at the PDF fit stage. To do this, we use the fact that a PDF fit may be recast in a physical basis, where the PDFs themselves are bypassed entirely, and one instead relates measured observables to predicted ones. In the case of factorization scale variation we find that in various situations there is a high degree of effective correlation between the variation in the fit and in predicted observables. In particular, including such a variation in both cases can lead to an exaggerated theoretical uncertainty. More generally, a careful treatment of this correlation appears mandatory, at least within the standard scale variation paradigm. For the renormalization scale, the situation is less straightforward, but again we highlight the potential for correlations between related processes in the fit and predictions to enter at the same level as between processes in the fit or prediction alone.

\end{abstract}
                                   
\end{center}

\section{Introduction}

The history of the determination of parton distribution functions (PDFs) 
from comparison to data goes back many decades, see~\cite{Gao:2017yyd} for a recent review. 
For some years the precision in both the data and theory was such that no systematic uncertainty estimate on the PDF at all was warranted or required. If some estimate of uncertainty 
were needed, then a comparison of different PDFs from different groups, or using 
different assumptions for one group, were thought to be sufficient.  The situation changed in the first years of the new millennium. This was largely driven 
 by the very precise measurements of structure function data over 
a wide range of both $x$ and $Q^2$ by the HERA experiment (see~\cite{Abramowicz:2015mha} for the final Run I + II combination).  
In addition, various apparent observed excesses over the Standard Model predictions, such as in high $E_\perp$ inclusive jet production at CDF~\cite{Abe:1996wy} were subsequently 
explained by a suitable modification of the PDFs~\cite{Huston:1995tw}, rather than being due to new physics. A systematic evaluation of PDF uncertainties therefore became 
essential, and the first global PDFs with an estimate of the uncertainty due to the experimental precision were released by CTEQ~\cite{Pumplin:2001ct,Pumplin:2002vw} and
MRST~\cite{Martin:2002aw}, building on earlier DIS--only fits~\cite{Alekhin:1996za,Botje:1999dj,Barone:1999yv,Giele:2001mr}.

In these first fits PDF uncertainties were 
a few percent at best, and much larger for many PDF flavours and $x$ regions. Soon after this
the full calculation of the next-to-next-to leading order (NNLO) splitting
functions for the evolution of PDFs were presented in full~\cite{mvvns,Vogt:2004mw} and 
NNLO extractions of PDFs became possible, as in e.g. \cite{Martin:2007bv,Alekhin:2009ni} (with minor approximations
for some data sets). At this point it was assumed that the 
theoretical precision on the PDFs was rather better than the uncertainties 
from the data, as well as being more difficult to quantify. Hence 
PDF uncertainties were always interpreted as being an experimental uncertainty due 
to the statistical and systematic uncertainties of the data included in the fit.  In recent years the 
understanding of the experimental uncertainties on PDFs has improved, and
as well as the Hessian-based approach of the original PDF fitters an 
alternative approach based on neural networks and generation of statistically
distributed replicas of PDFs has reached full maturity, as implemented in the NNPDF sets (see~\cite{Ball:2017nwa} for the most recent fit).
Good agreement between the two approaches, both for central values and uncertainties is seen~\cite{Butterworth:2015oua}. Indeed PDFs can be combined and those generated via the 
Hessian procedure can be converted to replicas and {\it vice versa} \cite{Watt:2012tq,Carrazza:2015aoa}. 

Any systematic consideration of 
theoretical uncertainties in PDF fits was limited to variations
due to changes in the strong coupling $\alpha_S(M_Z^2)$, quark masses and 
sometimes possible higher twist terms.
In addition to this, the 
convergence of different variants of variable flavour number schemes has
been demonstrated \cite{Andersen:2014efa}, and some agreement on the influence of use of 
variable flavour as opposed to fixed flavour schemes, and on fitting rather 
than imposing cuts for higher twist effects has been demonstrated 
\cite{Thorne:2012az,Ball:2013gsa,Thorne:2014toa}. However, these omit a potentially significant source of uncertainty,
 namely due to the fact that fixed--order perturbative predictions are used for the theoretical input to PDF fits. The uncertainty due to the approximate form of these, namely due to missing higher order (MHO) corrections, has until very recently not been considered at all, even on a semi--quantitative basis. 
We should note that introduction of a tolerance in PDF fits, such as the dynamic tolerance procedure for $\Delta \chi^2$ determination introduced in~\cite{Martin:2009iq}, is largely introduced to take account of tension between data sets in the fit, as justified in for example~\cite{Watt:2012tq}. However, 
some of the apparent tension between data sets in a fixed order fit may in fact be due to these missing higher order corrections. Such an effect is clearly seen in a LO PDF fit, for example, where the NLO corrections to DIS and Drell Yan cross sections are very different. Hence some part of the 
tolerance can likely be attributed to this, although in a NNLO fit this is probably a small component, and it is certainly challenging to quantify this precisely.
Nonetheless, it would be interesting to pursue a quantitative study of the degree 
to which observed tensions between data sets in the fit reduce as additional higher order corrections and/or uncertainties due to the missing higher orders are included.  

A more focussed study of the theoretical uncertainties in PDFs has just begun
in some quarters, and preliminary results have been presented in~\cite{Pearson:2018tim}. 
This is based on a variation of factorization and renormalization scales by a fixed factor of two in the theory input for the fit, a method that is 
frequently taken as the standard means of estimating MHO corrections in QCD. The purpose of this article is to critically examine a potentially important and quite general issue with taking such an approach, based on the fact  
 that the PDFs are not themselves physical quantities. In particular, we argue that this type of straightforward scale variation  does not provide a particularly obvious definition of what one means by `theoretical uncertainty' for PDFs. 
 
 In more detail, in practice one obtains the PDFs by fitting to data 
for one physical quantity (or more generally a set of them), and then predicts another physical quantity from 
these, using perturbatively calculated partonic cross sections for these quantities. Ultimately it is the uncertainty on the predicted 
quantity that is required. This clearly has a contribution from the 
experimental uncertainty on the PDFs, and this is included as standard. 
There is then also a theoretical uncertainty on the prediction arising 
from the finite order of the calculation for both the predicted cross section
and for the cross sections entering the PDF extraction. The former of these is 
included as standard (normally using scale variations) while the latter is 
not. However, we will argue in this article that for factorisation scale
variation there is a highly non-trivial interplay between the scale variation
when obtaining the PDF and when obtaining the required prediction, and this 
can potentially lead to a misinterpretation of the `theory uncertainty' on 
the prediction. In short, the aim of the process is to measure one 
physical quantity, and in terms of this predict another quantity. If, as seems natural, one
interprets the `theory uncertainty' as that inherent in 
expressing the predicted quantity in terms of the measured quantity due to 
MHOs in the relationship between the two, then we will 
show that varying the factorisation scale by a set amount in both the PDF
extraction and also in the prediction in terms of PDFs can lead 
to an effectively exaggerated factorisation scale variation when 
determining the full theoretical uncertainty. Our arguments rely on the fact that it is possible, and sometimes preferable (in principle at least), to bypass the intermediate PDFs entirely, instead working purely at the level of physical observables (structure functions and so on) and the relationships between them. This was behind the
original proposal of the DIS factorisation scale~\cite{Altarelli:1979ub}, and has subsequently been developed in for example\cite{Grunberg:1982fw,Catani:1996sc,Thorne:1997mb,Blumlein:2000wh,Hentschinski:2013zaa,Davies:2017hyl} under the general name of `physical schemes'.  

We also consider the case of the renormalization scale variation, which we find to be less amenable to this treatment, implying that the conventional approach should be reliable here. Nonetheless, one basic implication of working in this physical basis where one considers a PDF fit to be a (complex) relation between physical observables does follow in this case. Namely, any correlation between renormalization scale variations that one assumes is present in related physical processes entering the fit, should also in principle be included at the same level between fit and predicted processes. 

The outline of this paper is as follows. In Section~\ref{sec:qns} we consider the simplest possible case of fitting to a non--singlet structure function observable, and then predicting a second such structure function, and a non--singlet Drell--Yan cross section, in Sections~\ref{sec:qns1} and~\ref{sec:nsdy}, respectively. In Section~\ref{sec:gen} we generalise this to the case of structure functions involving both quark and gluon contributions, and comment on the corresponding high and low $x$ limits. In Section~\ref{sec:ren} we discus the case of renormalization scale variation. In Section~\ref{sec:conc} we discuss the implications of our findings and conclude. In Appendix~\ref{sec:ap1} we briefly summarise the case of DGLAP evolution in the diagonal basis at NLO.

\section{A simple example: the non--singlet quark}\label{sec:qns}

\subsection{Structure functions}\label{sec:qns1}

To illustrate the key issue, we start with the simplified case of a fit to a non--singlet structure function, and corresponding prediction of a second (distinct) non--singlet structure function. This represents the simplest example of our general argument, and having done this we will show how the case of a predicted hadronic observables, namely the non--singlet Drell--Yan cross section, follows straightforwardly. 

The structure function we consider is given purely in terms of the quark non--singlet distribution. A concrete example of this is the neutral current structure function $F_3^{\rm NC}$ due to $Z$ exchange, but we will for the sake of generality refer to the observable as $F_{{\rm NS}}$. We write this to NLO as
\be\label{eq:fns}
F_{{\rm NS}}(x,Q^2) = xq_{\rm NS}(x,\mu^2) + \tilde{\alpha}_S C_q^{(1)} \otimes xq_{\rm NS}(x,\mu^2) + \tilde{\alpha}_S \ln\left(\frac{Q^2}{\mu^2}\right) P_{qq}^{(0)} \otimes x q_{\rm NS}(x,\mu^2)\;,
\ee
where $\tilde{\alpha}_S\equiv \alpha_S/2\pi$, $\mu$ is the factorization scale, $C_q^{(n)}$ is the order $n$ coefficient function, and we consider for simplicity only one quark flavour. Note that for now we assume a fixed renormalization scale, the argument of which is suppressed for simplicity, and do not consider any issues related to its variation\footnote{We are implicitly using the ``standard'' convention, see e.g. \cite{SF}, that in the PDF evolution 
the scale of the coupling is taken to be the same as the factorization scale, 
i.e the PDFs depends on only one scale. However, the arguments all remain 
the same if this scale of the coupling in the evolution is related to the 
factorization scale by $\mu_R=c\mu_F$ if $c$ is the same for all 
physical quantities, i.e. the scale choice in the coupling for PDF evolution is not process--dependent.}. To keep the expressions which follow simpler we have also taken the leading order coefficient function as $C_q^{(0)}=1$, which can always be achieved by a suitable redefinition of the normalisation of $F_{\rm NS}$. We use the shorthand
\be
g\otimes f(x)  = \int_x^1 \frac{{\rm d}z}{z} f(z) g\left(\frac{x}{z}\right)\;,
\ee
throughout. The non--singlet quark combination $q_{\rm NS} = q-\overline{q}$ obeys the usual NLO DGLAP evolution
\be\label{eq:qdglap}
\frac{\partial q_{\rm NS}(x,\mu^2)}{\partial \ln\mu^2} = \tilde{\alpha}_S  \left(P_{qq}^{(0)}+\tilde{\alpha}_SP_{qq}^{(1)}\right)\otimes q_{\rm NS}(x,\mu^2)\;.
\ee
We now consider an idealised PDF fit of $q_{\rm NS}$ to the structure function, that is we assume that $F_{\rm NS}$ has been measured to arbitrary accuracy over the $x$ region we are interested in. We are free to do this as we are only considering the impact of theoretical uncertainties on the fit, and the inclusion of the (unrelated) experimental sources of uncertainty will not qualitatively effect the argument which follows. Indeed, we are precisely most interested in the case where the former dominates over the latter. Defining the ratio  $a_i = \mu^2/Q^2$, we can rewrite \eqref{eq:fns} as 
\begin{align}
xq_{{\rm NS}}(x,\mu^2) &= F_{{\rm NS}}(x,\mu^2/a_i) -  \tilde{\alpha}_S C_q^{(1)} \otimes xq_{{\rm NS}}(x,\mu^2)+\tilde{\alpha}_S  \ln a_i P_{qq}^{(0)} \otimes xq_{{\rm NS}}(x,\mu^2)\;,\\ \label{eq:qns2}
&=F_{{\rm NS}}(x,\mu^2/a_i) -  \tilde{\alpha}_S C_q^{(1)} \otimes  F_{{\rm NS}}(x,\mu^2/a_i)+\tilde{\alpha}_S  \ln a_i P_{qq}^{(0)} \otimes  F_{{\rm NS}}(x,\mu^2/a_i) \;,
\end{align}
where in the second line we consistently drop terms of $O(\alpha_S^2)$. Note a `standard' fit, i.e. where one does not consider any scale variation and takes the conventional choice of $\mu^2=Q^2$, simply corresponds to
\be\label{eq:qns1}
xq_{{\rm NS}}(x,\mu^2) =F_{{\rm NS}}(x,\mu^2) -  \tilde{\alpha}_S C_q^{(1)} \otimes  F_{{\rm NS}}(x,\mu^2)\;,
\ee
while for example a standard factor of 2 scale variation about the central scale $\mu = Q$ would correspond to taking $a_i \in (\frac{1}{4},4)$. 

We now use this to predict a second, distinct, non--singlet structure function, $F_{\rm NS}^\prime$:
\begin{align}\nonumber
F_{{\rm NS}}^\prime(x,Q^2) &= xq_{\rm NS}(x,\mu^2) + \tilde{\alpha}_S C_q^{\prime(1)} \otimes xq_{\rm NS}(x,\mu^2) + \tilde{\alpha}_S \ln\left(\frac{Q^2}{\mu^2}\right) P_{qq}^{(0)} \otimes x q_{\rm NS}(x,\mu^2)\;,\\\label{eq:fnst}
& = xq_{\rm NS}(x,a_f Q^2) + \tilde{\alpha}_S C_q^{\prime(1)} \otimes xq_{\rm NS}(x,a_f Q^2) - \tilde{\alpha}_S \ln a_fP_{qq}^{(0)} \otimes x q_{\rm NS}(x,a_f Q^2)\;,
\end{align}
where we have defined $a_f = \mu^2/Q^2$. We now consider the standard fit approach, that is using \eqref{eq:qns1} to express the quark distribution in terms of the structure function, giving 
\be\label{eq:f2fit}
F_{{\rm NS}}^\prime(x,Q^2) =F_{{\rm NS}}(x,a_fQ^2)  + \tilde{\alpha}_S \left(C_q^{\prime(1)} -C_q^{(1)} \right)\otimes  F_{{\rm NS}}(x,a_fQ^2) - \tilde{\alpha}_S \ln a_fP_{qq}^{(0)} \otimes x F_{\rm NS}(x,a_f Q^2)\;,
\ee
Thus we can rewrite our prediction so that no reference is made to the intermediate PDFs, and instead we express one observable quantity, $F_{\rm NS}^\prime$, in terms of the another, $F_{\rm NS}$. This expression is accurate to $O(\alpha_S)$ and following the standard rule of thumb approach we can then evaluate the theoretical uncertainty from MHOs by performing a scale variation with $a_f \in (\frac{1}{4},4)$.

We now consider the case that the scale is allowed to vary in the fit as well, 
which simply corresponds to keeping the $a_i$ dependence as in \eqref{eq:qns2}. We find
\be\label{eq:f2fit1}
F_{{\rm NS}}^\prime(x,Q^2) = F_{{\rm NS}}(x,a_{fi}Q^2)  + \tilde{\alpha}_S 
\left(C_q^{\prime(1)} -C_q^{(1)} \right)\otimes  F_{{\rm NS}}(x,a_{fi}Q^2) - 
\tilde{\alpha}_S \ln a_{fi} P_{qq}^{(0)} \otimes x F_{\rm NS}(x,a_{fi} Q^2)\;,
\ee
where we have defined $a_{fi}\equiv a_f/a_i$. Thus the $a_f$ and $a_i$ dependence is entirely contained 
in the ratio $a_f/a_i$, such that \eqref{eq:f2fit1} is  {\it identical} to \eqref{eq:f2fit} upon the replacement $a_f \to a_{fi}$.
Of course, if we expand out about e.g. fixed scale $Q^2$, then
\be
F_{{\rm NS}}(x,a_{fi}Q^2) = F_{{\rm NS}}(x,Q^2) + \tilde{\alpha}_S \ln a_{fi} 
P_{qq}^{(0)} \otimes x F_{\rm NS}(x,Q^2),
\ee
and the dependence on $\ln(a_{fi})$ in \eqref{eq:f2fit1} vanishes
at $O (\alpha_S)$ due to cancellations between the first and third terms, 
and only appears at $O(\alpha_S^2)$, where higher-order terms beyond
those shown in \eqref{eq:f2fit1} would be required for further cancellations.
We note also, that at this stage the distinction between $a_{fi}$ and $a_f$ is 
really entirely artificial. There is only one physical relation between the two
structure functions
\be\label{eq:f2fit2}
F_{{\rm NS}}^\prime(x,Q^2) =F_{{\rm NS}}(x,a Q^2)  + \tilde{\alpha}_S 
\left(C_q^{\prime(1)} -C_q^{(1)} \right)\otimes  F_{{\rm NS}}(x,a Q^2) - 
\tilde{\alpha}_S \ln a P_{qq}^{(0)} \otimes x F_{\rm NS}(x,a Q^2)\;,
\ee
where $a$ corresponds to the relative difference in scale at which we evaluate 
$F_{\rm NS}$ and $F_{\rm NS}^\prime$. In other words, we can see that the effect 
of varying the scale in the fit \eqref{eq:qns2} is the same as the previous 
approach, but with a larger range of variation, $a=a_{fi}\in (\frac{1}{16},16)$.

How should we interpret this result? The rule of thumb variation is applied to a broad category of observables, under the expectation that this will provide an estimate of the MHO uncertainty.  Concretely, one varies the logarithms in $a$ within a reasonable range, in order to track of the decreasing dependence on these with increasing perturbative order, but nonetheless keeping the argument $a$ to be $O(1)$ in order to avoid spoiling the overall perturbative convergence. The precise choice of $a \in (\frac{1}{4},4)$ above is of course arbitrary, but is nonetheless guided by these principles. 

We have seen in the above scenario that the intermediate PDFs themselves can be bypassed entirely in favour of a straightforward and arguably more fundamental relation between the physical observables $F_{\rm NS}$ and $F_{\rm NS}^\prime$. This simply reflects the fact that the PDFs are not themselves observables, and follows in a similar way to the physical factorization approach discussed elsewhere~\cite{Grunberg:1982fw,Catani:1996sc,Thorne:1997mb,Davies:2017hyl}. In terms of this relation, there is only one degree of freedom for scale variation, namely $a$ in \eqref{eq:f2fit2}. Within the context of the standard rule of thumb variation, the only reasonable and consistent choice appears to be to take \eqref{eq:f2fit2} and vary $a \in (\frac{1}{4},4)$. 

Now of course from a practical point of view one will not in general work explicitly in this physical framework, but rather in terms of the PDFs. The aim should therefore be to be consistent with the above results when doing so and evaluating a theoretical uncertainty on the PDFs themselves. We have seen above that in our example one should either vary the factorization scale in the prediction by the canonical factor of 2, or equivalently in the fit, but not in both. One may clearly call into question the reliability of such simple scale variations, but nonetheless at least under the assumption that this $a \in (\frac{1}{4},4)$ variation provides an accurate estimate of the theoretical uncertainty for general observables, this result will hold.

While the above result is demonstrated at NLO for simplicity, this remains true at arbitrary order.  This becomes clearest when we work in Mellin space, where we can write the DGLAP evolution \eqref{eq:qdglap} in the simple form
\be\label{eq:qnsmell}
q_{\rm NS}(j,\mu^2)=q_{\rm NS}(j,Q^2)\left(\frac{\mu^2}{Q^2}\right)^{\tilde{\alpha}_s\gamma_{qq} (j,\tilde{\alpha}_s)}\;,
\ee
where $j$ denotes the Mellin moment. Then, our expression for $F_{\rm NS}$, at a scale $Q$, can be written in the form
\begin{align}\nonumber
F_{\rm NS}(j,Q^2)&= q_{\rm NS}(j+1,\mu^2) \left(\frac{Q^2}{\mu^2}\right)^{\tilde{\alpha}_s\gamma_{qq}(j,\tilde{\alpha}_s)} c_q (j,\tilde{\alpha}_s)\;,\\ \label{eq:fnsmell}
&=q_{\rm NS}(j+1,\mu^2) a_i^{-\tilde{\alpha}_s\gamma_{qq}(j,\tilde{\alpha}_s)}c_q (j,\tilde{\alpha}_s)\;,
\end{align}
where  we define $a_i=\mu^2/Q^2$ as before. In the above expression, and in what follows, it is understood that the result at any arbitrary order in perturbation theory is given by expanding this out to the desired order in $\alpha_S$. Here
\be
\gamma_{qq}(j,\tilde{\alpha}_s)= \sum_{k=0}^n  (\tilde{\alpha}_s)^k \gamma_{qq}^{(k)}(j)\;,\qquad c_q (j,\tilde{\alpha}_s)=\sum_{k=0}^n (\tilde{\alpha}_s)^k c_q^{(k)}(j)\;,
\ee
correspond to the non--singlet $qq$ anomalous dimension and Mellin transform of the coefficient function at order $n$, respectively. Note that we have defined $c_q^{(0)}\equiv 1$ in the structure function case above, but we leave the expression completely general here. Now if we consider the second structure function at the same scale
$Q$, we have
\be
F^\prime_{\rm NS}(j,Q^2) = q_{\rm NS}(j+1,\mu^2) a_i^{\tilde{\alpha}_s\gamma_{qq}(j,\tilde{\alpha}_s)}c_q^\prime (j,\tilde{\alpha}_s)\;,
\ee
and so expressing $F^\prime_{\rm NS}(j,Q^2)$ in terms of $F_{\rm NS}(j,Q^2)$ we obtain the very simple result
\be
F^\prime_{\rm NS}(j,Q^2) = \frac{c_q^\prime (j,\tilde{\alpha}_s)}{c_q (j,\tilde{\alpha}_s)}F_{\rm NS}(j,Q^2)\;.
\ee
We can see that there is now no explicit dependence on the factorization scale, $\mu$, at all. The above situation is however defined for the quite specific case that we wish to relate the two observables at exactly the same scale.
More generally, we have the freedom to express $F^\prime_{\rm NS}(j-1,Q_f^2)$ in terms of $F_{\rm NS}(j-1,Q_i^2)$, that is
choose to express the predicted quantity a different physical scale, $Q_f$, to the 
scale at which we express the measured quantity, $Q_i$. In this case we have 
\be
F^\prime_{\rm NS}(j,Q_f^2) = \frac{c_q^\prime (j,\tilde{\alpha}_s)}{c_q(j,\tilde{\alpha}_s)}
\left(\frac{Q^2_f}{Q_i^2}\right)^{\tilde{\alpha}_s\gamma_{qq}(j,\tilde{\alpha}_s)}F_{\rm NS}(j,Q^2_i)\;.
\ee
Hence, we see that to any arbitrary order the relationship between the prediction and 
the measurement can be expressed as a ratio of scales, and that the 
relationship between the factorization scale chosen when fitting the PDF
to that chosen when making the prediction becomes equivalent to the 
relationship between the scale at which one predicts the physical quantity 
and that at which one evaluates the physical quantity entering the fit. To be completely explicit, we can suggestively define $Q^2 \equiv Q_f^2$ and $\mu^2 \equiv Q_i^2$, in terms of which we have
\be
F^\prime_{\rm NS}(j,Q^2) = \frac{c_q^\prime (j,\tilde{\alpha}_s)}{c_q(j,\tilde{\alpha}_s)}
\left(\frac{Q^2}{\mu^2}\right)^{\tilde{\alpha}_s\gamma_{qq}(j,\tilde{\alpha}_s)}F_{\rm NS}(j,\mu^2)\;,
\ee
where $F_{\rm NS}$ obeys the same DGLAP evolution equation \eqref{eq:qnsmell} as $q_{\rm NS}$, up to terms in the ratio of the coefficient function which depend 
on the running of the coupling; indeed, if we more correctly re--introduce the scale dependence of $\alpha_S$ into the above results, these would follow in precisely the same way, up to this difference in evolution.  Thus, we can express this at any arbitrary order, as is done in \eqref{eq:fns} at NLO, and we have an analogous freedom to vary the scale $\mu$ at which one evaluates $F_{\rm NS}$ as in the case of the standard factorization scale. However, as we have only one ratio of scales involved in this relation, there is only one such degree of freedom, and not the two implied by varying the factorization scale independently in the fit and prediction.

Finally, we note that while it might seem most direct in the above expression to choose the scale of $F_{\rm NS}$ ($\mu=Q_i$)
equal to the scale at which the measurement is made, this is, of course, not mandatory. If the scale 
at which one structure function is fit is significantly different to that at 
which the second is to be predicted ($Q=Q_f$) it would normally be assumed to be more 
sensible to express the measured quantity at a scale similar to the predicted 
quantity, relying on the validity of the evolution equation and avoiding 
obvious large logarithms in the expression relating the two physical 
quantities.  

\subsection{Drell--Yan Cross Section}\label{sec:nsdy}

As a second simple example of the above argument, and to demonstrate how the above result can readily be generalised to the case of hadronic observables, we can  use the same fit as above to predict the non--singlet Drell--Yan production cross section, i.e. $q_{NS}(x_1)q_{NS}(x_2) \to l^+ l^-$. We can write this at NLO as
\begin{align}\nonumber
\frac{{\rm d}\sigma^{\rm DY}_{\rm NS}}{{\rm d} Q^2} &= \int_0^1 {\rm d}x_1{\rm d}x_2 {\rm d}z\,\delta(x_1 x_2 z -\tau) q_{\rm NS}(x_1,\mu^2) q_{\rm NS}(x_2, \mu^2) \\
&\cdot\left[ \delta(1-z) + 2\tilde{\alpha}_S \left(P_{qq}^0(z) \ln \frac{Q^2}{\mu^2} + C_{{\rm DY},q}^{(1)}(z)\right)\right]\;,
\end{align}
where $Q^2$ is the dilepton invariant mass and $\tau = Q^2/s$. As before, for simplicity we normalise our cross section so that the LO coefficient function is unity. Defining $a_f =\mu^2/Q^2$, to the order we are calculating we can rewrite this as
\begin{align}\nonumber
\frac{{\rm d}\sigma^{\rm DY}_{\rm NS}}{{\rm d} Q^2} &=\int_\tau^1 \frac{{\rm d}x_1}{x_1}\left(q_{{\rm NS}}(x_1,a_f Q^2) - \tilde{\alpha}_S  \ln a_f  P_{qq}^{0} \otimes q_{{\rm NS}}(x_1,a_f Q^2)+  C_{{\rm DY},q}^{(1)}\otimes q_{{\rm NS}}(x_1,a_f Q^2) \right)\\ \nonumber
&\cdot \left(q_{{\rm NS}}(x_2,a_f Q^2) - \tilde{\alpha}_S   \ln a_f  P_{qq}^{0} \otimes q_{{\rm NS}}(x_2,a_f Q^2)+  C_{{\rm DY},q}^{(1)}\otimes q_{{\rm NS}}(x_2,a_f Q^2)\right)\;.
\end{align}
where $x_2=\tau/x_1$. Proceeding as before, and expressing the quark distribution in terms of the non--singlet structure function, we have 
\begin{align}\nonumber
\frac{{\rm d}\sigma^{\rm DY}_{\rm NS}}{{\rm d} Q^2} & =  \int_\tau^1 \frac{{\rm d}x_1}{x_1} \Big(F_{\rm NS}(x_1,a Q^2) - \tilde{\alpha}_S C_q^1 \otimes F_{\rm NS}(x_1,a Q^2)- \tilde{\alpha}_S  \ln a  P_{qq}^{0} \otimes F_{\rm NS}(x_1,a Q^2)\\ \nonumber
&+  C_{{\rm DY},q}^{(1)}\otimes F_{{\rm NS}}(x_1,a Q^2)\Big) \Big(F_{\rm NS}(x_2,a Q^2)- \tilde{\alpha}_S C_q^1 \otimes F_{\rm NS}(x_1,a Q^2) \\ \label{eq:ffit}
&- \tilde{\alpha}_S   \ln a  P_{qq}^{0} \otimes F_{\rm NS}(x_2,a Q^2)+  C_{{\rm DY},q}^{(1)}\otimes F_{{\rm NS}}(x_2,a Q^2)\Big)\;,
\end{align}
where $a=a_f$ for the case that we vary the scale only in the prediction, and $a=a_{fi}$ if we vary in both cases. In other words, the argument follows through in exactly the same way.

Exactly as in the previous example, we note that the above results can be more simply expressed in Mellin space. In particular, taking the Mellin transform with respect to $\tau$ the DY cross section simply becomes
\begin{align}\nonumber
\frac{{\rm d}\sigma^{\rm DY}_{\rm NS}(j,Q^2)}{{\rm d} Q^2} &= \left(q_{\rm NS}(j,\mu^2)a_f^{\tilde{\alpha}_s\gamma_{qq}(j,\tilde{\alpha}_s)}\right)^2 \left[1+2\tilde{\alpha}_S  c_q^{\rm DY}(j)+ \cdots\right]\;,\\
&= \left(F_{\rm NS}(j-1,a_{fi}Q^2) a_{fi}^{\tilde{\alpha}_s\gamma_{qq}(j,\tilde{\alpha}_s)}\right)^2\left[1+2\tilde{\alpha}_S  c_q^{\rm DY}(j)+ \cdots\right]\;,
\end{align}
where the anomalous dimension can be calculated at any arbitrary order, and the `$\cdots$' indicate these higher order contributions to the coefficient function. 
\section{A more general example: the quark singlet and gluon}\label{sec:gen}

\subsection{Set-up}

The above examples considered the special case of observables given in terms of a single non--singlet quark distribution. This leads to a simple and transparent result, but it is not immediately clear how it will generalise to the case that includes both quark and gluon partons, which obey the fully coupled DGLAP equation. 

For the sake of generality and demonstration, we consider a PDF fit to a pair of arbitrary `structure function' observables $F$ and $H$, which are then used to predict a third such observable, $K$. The generalisation to the case of hadronic observables would render the corresponding analysis a great deal more complex in practice, but in principle should not change the basic argument. We will also work in Mellin space, as this will simplify the calculation, although all the results which follow hold analogously in $x$ space as well. We write
\begin{align}\nonumber
F(j,Q^2)& = \Sigma_q(j+1,\mu^2) F_q \left(j,\frac{Q^2}{\mu^2}\right)+ g(j+1,\mu^2) F_g \left(j,\frac{Q^2}{\mu^2}\right)\;,\\ \label{eq:fh}
H(j,Q^2)&= \Sigma_q(j+1,\mu^2) H_q \left(j,\frac{Q^2}{\mu^2}\right) + g(j+1,\mu^2) H_g \left(j,\frac{Q^2}{\mu^2}\right)\;.
\end{align}
Thus, we assume that these observables depend on the gluon ($g$) and total quark singlet ($\Sigma_q$) PDFs only. In other words, any dependence on non--singlet quark combinations, which would be introduced by e.g. including a quark flavour--dependence (due typically to the quark EW charges), is omitted to limit the observables we need to consider, although the set--up can readily be generalised to include these.

We will drop the Mellin moment argument $j$ for brevity in what follows. We can write the coefficient functions at NLO as
\begin{align}
F_q \left(\frac{Q^2}{\mu^2}\right) & = c_q^{(0)} + \tilde{\alpha}_S c_q^{(1)} + \tilde{\alpha}_S c_q^{(0)} \ln\left(\frac{Q^2}{\mu^2}\right) \gamma_{qq}^{(0)}  + \tilde{\alpha}_S c_g^{(0)} \ln\left(\frac{Q^2}{\mu^2}\right)\gamma_{gq}^{(0)} \;,\\
F_g \left(\frac{Q^2}{\mu^2}\right) & = c_g^{(0)} + \tilde{\alpha}_S c_g^{(1)} + \tilde{\alpha}_S c_q^{(0)} \ln\left(\frac{Q^2}{\mu^2}\right) \gamma_{qg}^{(0)}  + \tilde{\alpha}_S c_g^{(0)} \ln\left(\frac{Q^2}{\mu^2}\right)\gamma_{gg}^{(0)} \;,
\end{align}
and similarly for $H$. Note that the $q\leftrightarrow g$ mixing introduces a corresponding mixing in the coefficients $c_{q,g}$ of the expansions of the $F_{q,g}$, and similarly for $H_{q,g}$. This simplifies if we instead use the basis of eigenvectors of the DGLAP equation, which we denote $\Sigma_\pm$. In terms of these we can write
\begin{align}
F(Q^2)&= \Sigma_+ (\mu^2) \left(\frac{Q^2}{\mu^2}\right)^{\tilde{\alpha}_S \gamma_+} F_+ + \Sigma_- (\mu^2) \left(\frac{Q^2}{\mu^2}\right)^{\tilde{\alpha}_S \gamma_-} F_-\;,\\
H(Q^2) &=\Sigma_+ (\mu^2) \left(\frac{Q^2}{\mu^2}\right)^{\tilde{\alpha}_S \gamma_+} H_+  +\Sigma_- (\mu^2) \left(\frac{Q^2}{\mu^2}\right)^{\tilde{\alpha}_S \gamma_-} H_-\;,
\end{align}
at arbitrary order. Here the diagonal anomalous dimensions $\gamma_\pm$ and the PDF eigenvectors are given explicitly in Appendix~\ref{sec:ap1}. These then define the coefficients $F_\pm$ and $H_\pm$ above, which are given in terms of $F_q$ and $F_g$, and similarly for $H$.

We now consider  a PDF fit to these observables, for which we take the factorization scales $\mu^2 = a_{f} Q^2$ and $\mu^2 = a_{h} Q^2$. In this case we have (see \eqref{eq:sigev})
\begin{align}
F\left(\frac{\mu^2}{a_f}\right)&= \Sigma_+ (\mu^2) \left(a_f\right)^{-\tilde{\alpha}_S \gamma_+} F_+ + \Sigma_- (\mu^2) \left(a_f\right)^{-\tilde{\alpha}_S \gamma_-} F_-\;,\\
H \left(\frac{\mu^2}{a_h}\right)&=\Sigma_+ (\mu^2) \left(a_h\right)^{-\tilde{\alpha}_S \gamma_+} H_+  +\Sigma_- (\mu^2) \left(a_h\right)^{-\tilde{\alpha}_S \gamma_-} H_-\;,
\end{align}
which we can then invert to give
\begin{align}
\Sigma_+(\mu^2) &= \frac{H_- F\left(\frac{\mu^2}{a_f}\right) \left(a_f\right)^{\tilde{\alpha}_S \gamma_-} - F_- H \left(\frac{\mu^2}{a_h}\right) \left(a_h\right)^{\tilde{\alpha}_S \gamma_-}}{F_+ H_- \left(a_f\right)^{\tilde{\alpha}_S (\gamma_- - \gamma_+)}-F_- H_+  \left(a_h\right)^{ \tilde{\alpha}_S (\gamma_- - \gamma_+)}}\;,\\
\Sigma_-(\mu^2)&=\frac{H_+ F\left(\frac{\mu^2}{a_f}\right) \left(a_f\right)^{\tilde{\alpha}_S \gamma_+} - F_+ H \left(\frac{\mu^2}{a_h}\right) \left(a_h\right)^{\tilde{\alpha}_S \gamma_+}}{F_- H_+ \left(a_f\right)^{-\tilde{\alpha}_S (\gamma_- - \gamma_+)}-F_+ H_-  \left(a_h\right)^{- \tilde{\alpha}_S (\gamma_- - \gamma_+)}}\;.
\end{align}
Substituting these into the third structure function $K$ at scale $\mu^2 = a_{k} Q^2$, at NLO we find
\begin{align}\nonumber
K(Q^2) =&\frac{1}{F_+H_- - F_- H_+}\bigg\{K_+ H_- - H_+ K_- + \tilde{\alpha}_S \ln\left(\frac{a_f}{a_k}\right)\left[\gamma_+ K_+ H_- - \gamma_- K_- H_+\right]\\ \nonumber
& + \tilde{\alpha}_S (\gamma_- - \gamma_+)  \ln\left(\frac{a_h}{a_f}\right)\frac{H_+ H_-}{F_+ H_- - F_- H_+} (K_+ F_- - K_- F_+)\bigg\}F\left(\frac{a_k}{a_f} Q^2\right)\\ \nonumber
&+\frac{1}{H_+F_- - H_- F_+}\bigg\{K_+ F_- - F_+ K_- + \tilde{\alpha}_S \ln\left(\frac{a_h}{a_k}\right)\left[\gamma_+ K_+ F_- - \gamma_- K_- F_+\right]\\ \label{eq:fkf}
& - \tilde{\alpha}_S (\gamma_- - \gamma_+)  \ln\left(\frac{a_h}{a_f}\right)\frac{F_+ F_-}{H_+ F_- - H_- F_+} (K_+ H_- - K_- H_+)\bigg\}H\left(\frac{a_k}{a_h} Q^2\right)\;,
\end{align}
where it is understood that the fixed--order coefficients, $H_\pm$ and so on, should also be expanded out to the appropriate order, but we omit this for clarity. For comparison, if we do not vary the scales in the fit, i.e. we take $a_f = a_h=1$, then we have 
\begin{align}\nonumber
K(Q^2) =&F\left(a_k Q^2\right)\frac{1}{F_+H_- - F_- H_+}\bigg\{K_+ H_- - H_+ K_- - \tilde{\alpha}_S \ln a_k \left[\gamma_+ K_+ H_- - \gamma_- K_- H_+\right]\bigg\}\\ \label{eq:k}
&+H\left(a_k Q^2\right)\frac{1}{H_+F_- - H_- F_+}\bigg\{K_+ F_- - F_+ K_- - \tilde{\alpha}_S \ln a_k \left[\gamma_+ K_+ F_- - \gamma_- K_- F_+\right]\bigg\}\;.
\end{align}
Thus we can immediately see that the situation is somewhat more complex then in the simplified purely non--singlet case considered in the previous section. In particular, our general expression \eqref{eq:fkf} contains the two ratios $a_{f,h}/a_k$ corresponding to the arguments of the $F,H$, similar to the non--singlet case we considered before. However in addition we can see that the result includes a contribution that depends on the ratio $a_h/a_f$, which is purely due to the scale variation in the fit stage, and is completely absent in the prediction without this variation.
We note that while the above expression is written in terms of three ratios, only two of these are independent, exactly as we would expect following the discussion towards the end of Section~\ref{sec:qns1}. In particular,  we are now
expressing one predicted physical quantity defined at one scale in terms of 
two measured physical quantities, each of which may be evaluated at a different scale. 
There are therefore two independent scale ratios, and two physically meaningful
ratios of factorization scales. On the other hand, while the ratio  $a_h/a_f$ can be written in terms of the independent ratios $a_{f,h}/a_k$, we can see that this results in a mixing of these two ratios which does not immediately reduce to the simple situation we had in the case of the non--singlet structure functions.

We note that in~ \cite{Pearson:2018tim} it was advocated that, as all physical quantities share common PDFs, the factorization scale should be varied in a fully correlated way across all processes entering the fit. In the above analysis, we can see that this corresponds to taking $a_f=a_h$, and we are left with \eqref{eq:k}, after replacing $a_k \to a_k/a_f = a_f/a_h$. In other words, our situation and conclusions are exactly the same as for the simpler
non-singlet case, i.e. variations of factorization scale in the predictions
are entirely equivalent to those in the fitting and {\it vice versa}. We fully expect this to hold in the more general case appropriate to a global fit, as here too we will only have one independent ratio of scales for any given predicted process. Thus, if one makes this assumption, one could bypass the complication of including these variations at the fit stage entirely and simply include them in the prediction, with the assumption of full correlations implying that this should be done in the same way for different predictions at the same time. On the other hand, varying the factorization scale in both the fit and prediction would be a type of double counting, i.e varying the scale by a factor of two more than 
may naively be expected. 

However, as discussed further below, in general this appears to be an overly constraining assumption, given there is the question of the central choice of scale to consider and, potentially more significantly, the fact that the partons and $x$ range probed by the fit processes can be rather different. With this in mind, how do we interpret our above result if we do not make this simplifying assumption of fully correlated factorization scales for quantities in the PDF fit? We will first consider this result in various kinematic limits, before discussing the more general implications.

\subsection{The low and high $x$ limits}

One can simplify the full result \eqref{eq:fkf} by for example assuming that $F_- = 0$ and $H_+=0$, in other words that $F,H$ are only sensitive to the contribution from either the negative or positive eigenvectors. In this case we have
\be
K(Q^2) = \frac{K_+}{F_+}\left\{1+\tilde{\alpha}_S \ln\left(\frac{a_f}{a_k}\right) \gamma_+\right\}F\left(\frac{a_k}{a_f} Q^2\right)+\frac{K_-}{H_-}\left\{1+\tilde{\alpha}_S \ln\left(\frac{a_h}{a_k}\right)\gamma_-\right\}H\left(\frac{a_k}{a_h} Q^2\right)\;,
\ee
and the terms proportional to the ratio $a_f/a_h$ vanish.
As an aside, if in addition the prediction is only sensitive to one eigenvector, then this will reduce to a single ratio, in precisely the same way we saw for the non--singlet distribution.

While at first glance it may appear somewhat arbitrary to consider these limits, in fact in the low and high $x$ regions this can be precisely the situation we find ourselves in. 
As discussed further in Appendix~\ref{sec:ap1}, if we take the high $x$ limit, we have the well known result that
\be
\Sigma_+(j,\mu^2)  = g(j,\mu^2) \;, \qquad \Sigma_-(j,\mu^2) =  \Sigma_{q}(j,\mu^2)\;,
\ee
that is the quark distribution ($\sim \Sigma_q, q_{NS}$) and gluon are independent eigenvectors of the DGLAP evolution. In the alternative low $x$ ($j\sim 1$) limit we have
\be
g(j,\mu^2) \sim q(j,\mu^2) \sim \Sigma_+(j,\mu^2)\;,
\ee
for sufficiently high scale $\mu$. That is, the positive eigenvector is dominant.

These regimes play a direct role in PDF phenomenology at the LHC and elsewhere. For example, a topical case is the high $x$ gluon, which is relatively poorly determined, and in which there is currently a great deal of interest in placing further constraints. This typically involves the use of LHC observables such as inclusive jet and $t\overline{t}$ production, and  the $Z$ boson $p_\perp$ distribution, for which the high $x$ gluon plays a dominant role. Although a global fit includes of course a wider dataset, the extracted high $x$ gluon will to a significant extent be driven by these. One can then take the result of this fit and predict the gluon--initiated production of e.g. a high mass BSM object. In such a scenario the gluon evolution will be effectively decoupled and both the fit and predicted observables will be dominated by the positive eigenvector $\Sigma_+$. In other words, we are in an analogous situation to Section~\ref{sec:qns}, where the factorization scales for the fit and prediction are fully correlated, and varying the factorization scale in both will lead to and effective double counting of the theoretical uncertainty. The corresponding situation for the high $x$ quark, where both the singlet and non--singlet are decoupled from the gluon, is similar.

At low $x$, as we increase the scale of the observed process we find that the quark and gluon contribution are completely correlated by evolution, and only the positive eigenvector contributes. This is equally true for observables such as scaling violations of the structure function, ${\rm d} F_2/{\rm d}\ln Q^2$, which only depends on the positive eigenvector at low $x$, for all scales. Thus, for fit processes such as a ${\rm d} F_2/{\rm d}\ln Q^2$ and predicted processes such as  Drell--Yan production at the LHC (in particular in the lower mass region), we will expect a large degree of correlation.

We note that in a realistic PDF fit we will in general include multiple 
observables at the fit stage which may be dominated by a particular PDF 
eigenvector. This will therefore introduce a common set of factorization scale 
dependent logarithms in to the corresponding predictions, or equivalently the 
scale evolution of these observables will be same up to running coupling 
effects. It therefore seems natural in such a case to vary the factorization scale about some fixed
central value for each data set in a correlated way across these observables, either in the fit or 
prediction stage (but not both). However,
the choice of central/best fit is not obvious, e.g. $\mu=Q^2$ may be more appropriate
for DIS data and $\mu = M/2$ may be more appropriate for Drell Yan production. 
Further to this, while one might argue that if one varies the scale for one type of structure 
function from a central value of $\mu=Q$ to $\mu=2Q$, then one does it for all structure functions, it 
does not seem so clear that e.g. for some jet data related observable one should 
simultaneously vary the scale from a central value of e.g. $\mu=p_T$ to 
$\mu=2p_T$\footnote{Indeed, for some quantities the choice between $\mu=p_T$,
$\mu=p_{T,\max}$ or $\mu=\hat H_{T}$ is also open in principle, see e.g. 
\cite{Currie:2018xkj} for a recent discussion for the case of inclusive jets.
For dijets one has additionally has the choice of a $p_T$-based scale or the invariant mass
$m_{jj}$.}.  Such scale allocations are therefore not in any clear sense correlated, and certainly 
the degree of variation in the cross section when applying the rule of thumb variation of the 
scale will depend on the central scale, the precise `preferred' value of which is not necessarily clear.
We do not advocate strictly 
fitting the best scale for each type of process, but advise that some note, 
based on experience, should probably be taken to use central 
scales that provide good fits for a given process (or at the very least, to 
avoid those known not to be optimal). 

As described above, even for processes that depend dominantly on the same eigenvector, 
the correlation of the scale variation between these between processes is not entirely trivial. 
However,  if two quantities (either fit and predicted, or both fit) 
are dominated by alternative PDF eigenvectors and therefore completely 
independent, then clearly the variation of scales, in either fit or 
prediction, is not correlated. In this case imposing correlation between scale variations in these 
processes can result in artificial correlations between the predicted processes (depending on how they 
depend on the initial PDFs), or 
in the context of a fit, the PDFs themselves. In general, most predictions will depend 
on combinations of fit PDFs where full correlation is, to a lesser or greater 
degree, an overly restrictive assumption.

\section{Renormalization scale Variation}\label{sec:ren}

In this section we will discuss how the issues presented above for 
factorization scale variation apply to the separate question of renormalization 
scale variation. 
We  
define a physical quantity beginning at ${\cal O}(\alpha_S)$:
\be\label{eq:Ars}
A(Q^2) = \tilde\alpha_S(\mu_i^2)A_1 + \tilde{\alpha}_S^2(\mu_i^2) 
\left(A_2-\ln \beta^{(0)}A_1\left(\frac{Q^2}{\mu_i^2}\right) \right)\;,
\ee
where $\mu_i$ is the renormalization scale. We first consider this to be the quantity
we measure in order to determine the 
value of the coupling. With this, we wish 
to predict the second quantity
\be\label{eq:Brs}
B(Q^2) = \tilde\alpha_S(\mu_f^2)B_1 + \tilde{\alpha}_S^2(\mu_f^2) 
\left(B_2- \beta^{(0)}B_1\ln\left(\frac{Q^2}{\mu_f^2}\right)\right)\;.
\ee
As with the PDFs and structure functions one can invert \eqref{eq:Ars}
to obtain 
\begin{align}\nonumber
 \tilde\alpha_S(\mu^2)&=\frac{A(\mu^2/a_i)}{A_1}\left(1-\frac{\alpha_S(\mu^2)}{A_1}\left(A_2 +\beta^{(0)} A_1\ln a_i \right)\right)\;,\\
 & =\frac{A(\mu^2/a_i)}{A_1}\left(1-\frac{A(\mu^2/a_i)}{A_1^2}\left(A_2 +\beta^{(0)} A_1\ln a_i \right)\right)\;, \label{eq:invert}
\end{align}
where as usual $a_i=\mu_i^2/Q^2$ and we drop terms of ${\cal O}(\alpha_S^2)$; note that while the physical quantities begin at $O(\alpha_S)$, the relation between them is one order lower, and is accurate to $O(\alpha_S)$.
For $B(Q^2)$, taking $a_f=\mu_f^2/Q^2$ and substituting in our expression for $A(Q^2)$, we get
\be\label{eq:BrsA}
B(Q^2) = \frac{A(a_{fi} Q^2)}{A_1}\left(B_1 + \frac{A(a_{fi} Q^2)}{A_1}\left[B_2 - \frac{B_1 A_2}{A_1}+ \beta^{(0)} B_1 \ln a_{fi}\right]\right)\;,
\ee
Hence, we can indeed derive a final result which is expressed 
only in terms of a single ratio $a_{fi}=a_f/a_i$, as in the structure function case.

In the above example we have used our initial physical quantity in 
order to determine the coupling constant, not the PDFs. This is really the 
most natural thing to to when thinking about renormalization scale variation, 
since the scale is that used for the definition of the coupling, while the 
factorization scale is that used for the definition of the PDFs. However, in 
a PDF fit one instead uses the 
physical quantity $A(Q^2)$ to determine the PDFs. To see how this changes the result, we now assume our toy observables depend on a single PDF $q$ (the non--singlet quark, say, although this is not essential).
Implicitly we work in Mellin space to avoid complications with 
convolutions, but as before this does not change the basic argument.  We consider the case of a fixed factorization scale $\mu_F^2 = Q^2$, while setting $\mu_i^2 = a_i Q^2$ for the renormalization scale. We have 
\be\label{eq:Arspdf}
A(Q^2) = \tilde{\alpha}_S(\mu_i^2)A_1\left[1 + \frac{\tilde{\alpha}_S(\mu_i^2)}{A_1}
\left(A_2+ \beta^{(0)}A_1\ln a_i \right)\right]q(Q^2)\;.
\ee 
In terms of this the PDF can be written as
\be
q(Q^2) = \frac{A(Q^2)}{A_1 \tilde{\alpha}_S(\mu_i^2)}\left[1-\frac{\tilde{\alpha}_S(\mu_i^2)}{A_1}\left(A_2+ \beta^{(0)}A_1\ln a_i \right)\right]\;,
\ee
Inserting this into the expression for $B(Q^2)$, we obtain
\begin{align}\nonumber
B(Q^2) &= \tilde{\alpha}_S(\mu_f^2)B_1\left[1 + \frac{\tilde{\alpha}_S(\mu_f^2)}{B_1}
\left(B_2+ \beta^{(0)}B_1\ln a_f \right)\right]q(Q^2)\;\\ \label{eq:BrsApdf}
& = \frac{B_1 A(Q^2)}{A_1}\frac{\tilde{\alpha}_S(\mu_f^2)}{\tilde{\alpha}_S(\mu_i^2)}\left[1+\beta^{(0)}\left(\tilde{\alpha}_S(\mu_f^2)\ln a_f-\tilde{\alpha}_S(\mu_i^2)\ln a_i\right)+\frac{B_2}{B_1}\tilde{\alpha}_S(\mu_f^2)-\frac{A_2}{A_1}\tilde{\alpha}_S(\mu_i^2)\right]\;.
\end{align}
We see that now the expression is certainly not just a function of the ratio of 
scales $\mu_f^2/\mu^2= a_{fi}$. Let us examine the explicit consequence of this. 
For example, in our earlier case of factorisation scale variation, the choice of $\mu_f^2=\mu_i^2$ i.e. $a_{fi}=1$ resulted in no change in the 
expression for the prediction compared to choosing both $a_i=a_f=1$.  
In fact for the first term in \eqref{eq:BrsApdf} this equivalent result 
appears, and similarly for the sum of the second and third terms. However, for 
the fourth and fifth terms, i.e. those dependent on the coefficients of the 
scale-independent parts of the NLO expressions for $A(Q^2)$ and $B(Q^2)$, this 
is not the case. In this limit each becomes proportional to 
$\tilde \alpha_s(\mu_f^2)$, but depends on the absolute value of the scale.
This term depends on the difference in the relative size of these NLO 
corrections (compared to the LO contributions to each quantity), so the 
violation of the dependence on ratios is violated by the scale independent 
NLO corrections. If we consider other types of scale variation, e.g. 
multiplying $\mu_f$ by 2 but dividing $\mu_i$ by 2 we see that even though 
the effect in the combination of the first, second and third terms is close
to effect of either multiplying $\mu_f$ by 4 or dividing $\mu_i$ by 4, it 
is not identical, and the discrepancy is larger in the fourth and fifth terms.  

The fact that the expression of the predicted physical quantity in terms of 
the measured physical quantity does not break down into an expression depending 
on the ratio of the renormalization scales used for each calculation is a 
consequence of the fact that the renormalization scale is fundamentally 
associated to the scale of the coupling, but here we do not directly relate
the physical quantities to the coupling constant, but to the PDF. It is also the case that
different physical quantities depend on the coupling in different ways, i.e. 
the perturbative order starts at zeroth, first or second order for very 
standard quantities (and higher order for more exclusive quantities). Here we 
have given perhaps the simplest example of two quantities which each start at 
first order. However, the common input in PDF fits of the $F_{2,3}$ structure 
functions starts at zeroth order, so at lowest order has no renormalization 
scale dependence in the hard cross section. The renormalization scale dependence of $F_2$ will therefore be
suppressed by a power of $\alpha_S$ relative to the case of e.g. 
top-pair production in hadron-hadron scattering, which begins at 
${\cal O}(\alpha_S^2)$. 
 In contrast, all cross sections are linear in the PDF 
of any of the hadrons participating in the scattering.          

Finally, we can also consider the case of two related physical observables. This could be for example, jet production at ATLAS and CMS, or more generally $W$ and $Z$ boson production, for which the LO results are of course uncorrelated in normalization, but the effect of higher--order QCD corrections is  similar. Considering the latter example, for our toy observables above we would have
\be
\frac{A_2}{A_1} \sim \frac{B_2}{B_1} \equiv C_{\rm NLO}\;,
\ee
and so \eqref{eq:BrsApdf} can be written as
\be
\frac{B(Q^2)}{A(Q^2)}  = \frac{B_1}{A_1}\left[1+\beta^{(0)}\left(\tilde{\alpha}_S(\mu_f^2)\ln a_f-\tilde{\alpha}_S(\mu_i^2)\ln a_i\right)+C_{\rm NLO}\left(\tilde{\alpha}_S(\mu_f^2)-\tilde{\alpha}_S(\mu_i^2)\right)\right]\;.
\ee
Now to maintain consistency with our requirement that this ratio should be approximately constant under inclusion of higher--order QCD corrections, we can see that we must take $\mu_f = \mu_i$ ($a_f=a_i$), i.e. vary the renormalization scale in the fit and prediction in a correlated way. It is of course a well--known procedure to vary QCD renormalization scales in such a way when predicting this type of ratio (of e.g. the $W$ to $Z$ boson cross sections\footnote{See for example 
\cite{Lindert:2017olm} for the example of vector boson plus jets. In this study an additional, conservative, process dependent uncertainty is also introduced
to account for the difference between the $K$-factors of the different quantities.}). In~\cite{Pearson:2018tim} this argument is extended to hold between processes at the fit stage.  It is perhaps not particularly surprising to find an equivalent requirement between the fit and prediction here, and clearly the inclusion of this in a global fit would be intractable. Nonetheless, we can see that this correlation enters in principle at the same level as that between processes entering the fit, and so the question of whether it is necessary or sensible to include one without the other requires further investigation. Certainly, the relative importance of the correlation between processes in the fit stage and between the fit and prediction will in general depend on the specific data sets being considered.

\section{Summary and Conclusions}\label{sec:conc}

In this paper we have discussed the inclusion of theoretical uncertainties in PDFs due to  missing higher--order  terms in the pQCD results for the processes entering the fit. Such uncertainties, while routinely included in the predictions, have previously not been explicitly included in the PDF fit itself. We are now firmly in the high precision LHC era, both in terms of the available data for PDF fitting and the standard for phenomenology which applies these PDFs. Therefore, such an approach may be increasingly called into question, and certainly requires careful consideration.

As a first step towards this, we have considered the standard approach to evaluating MHO uncertainties, namely due to QCD factorization and renormalization scale variation around a central value by some set factor. Focussing on the case of the factorization scale, we have in particular shown that if we take this standard criterion seriously and apply it consistently to both a PDF fit and the predicted observables resulting from that fit, then in general there is a strong overlap between the variation in the fit and prediction stage. To demonstrate this, we have considered in Section~\ref{sec:qns} the simplest possible case of a fit to a non--singlet structure function, before generalising to include coupled quark and gluon contributions in Section~\ref{sec:gen}. We have shown how the explicit dependence on the PDFs can be removed entirely, and the outcome of the fit recast instead in terms of observable quantities only. We have then found that written in this way, scale variation in the fit corresponds to precisely a scale variation in the prediction in certain regimes, in particular at low or high enough $x$. Our results have relied on the basic fact that it is possible, and sometimes preferable, to bypass the intermediate PDFs entirely, instead working purely at the level of physical observables (structure functions and so on) and the relationships between them. This idea of working in such a `physical basis' is in fact quite an old one; here we simply derive the implications  of this for scale variation uncertainties in PDF fits.

We have also briefly considered the case of renormalization scale variation, finding that the situation is not as straightforward. This is unsurprising, given the quite different roles that the the QCD coupling and PDFs play in fits. However, one basic implication of working in a physical basis is that the motivation for including correlated renormalization scale variations between related processes in the fit or prediction state, is equally present between processes entering both the fit and prediction. While including such correlations in a realistic global fit would certainly in practice be impossible, clearly this raises questions if for example one wishes to include such correlations at the fit stage.

Now, the true situation in a global fit is certainly significantly more complicated than the examples we have considered explicitly in this paper. 
Here, we fit a very wide array of structure function and hadron collider data, sensitive to a range of different (and overlapping) $x$ values and scales. Indeed, while in the simple non--singlet case of Section~\ref{sec:qns} we find a complete overlap between the fit and prediction, we have seen that in the somewhat more general (although still simplified) scenario of Section~\ref{sec:gen} the situation is not as straightforward. Nonetheless, as mentioned above in certain (e.g. low and high $x$) regimes the same conclusion holds, and more generally these considerations serve as clear guidance for the case of a genuine PDF fit. In particular, a naive variation of factorization scales in both the PDF fit and prediction will certainly correspond to a degree of overestimation in the total theoretical error, and should be avoided. 
 On the other hand, the considerations of Section~\ref{sec:gen} also suggest that variation of the overall factorization scale in the prediction alone does not capture the full degree of uncertainty due to MHOs in the problem. This suggests that the correct approach, maintaining generality while avoiding overlap in the regimes where it may occur, would be to consider scale variations only in the fit and not in the prediction. 
 
Further to this, we have also seen that if one considers  
factorization scales between all quantities to be fully correlated in the fit, 
then the factorization scale variation can equivalently
be performed entirely in the calculation of the predictions, with the first 
assumption implicitly leading to full correlation also being maintained between the factorization scale across 
different predictions.  As discussed earlier, this seems to be a overly strong assumption in general, 
but as the arguments in Section 3 suggest it may, in practice, not be such a bad assumption for certain 
specific physical quantities. Therefore, for factorisation scale variations, the current approach of only 
varying the scale in the prediction is certainly an underestimate of the full uncertainty from this source, but probably
not as significant an underestimate as might naively be expected. The arguments in this article then suggest it is more reliable to consider  factorization scale in the fit alone, but this should in general include a variation which is only correlated for physical processes which depend on the same independent PDF
combinations. This type of procedure would clearly need 
significant compromise in practice, as few physical processes depend on exactly the same PDFs, so many sets of 
processes will either be weakly correlated, strongly correlated, or somewhere in between.
We note that in all cases the choices of central scale is largely a separate issue: this should be taken independent for different 
quantities, allowing for something close to the best possible fit at a fixed theoretical order, 
and possibly relieving some tensions between data sets in a fit. Indeed, such an approach also has the likely benefit of reducing the sensitivity of the fit quality, $\chi^2$, to MHOs, which may confuse the interpretation of PDF fits.

The eventual interpretation of these results has the potential to be a matter of some debate, given the known issues with the `rule of thumb' scale variation approach and availability of alternative, potentially superior, approaches (see for example~\cite{Cacciari:2011ze,David:2013gaa,Bagnaschi:2014wea}). Nonetheless the initial investigations of the inclusion of theoretical uncertainties in PDF fits currently apply the scale variation paradigm~\cite{Pearson:2018tim,Gao:2013xoa,SF}, and so this result is certainly directly relevant to these studies. Thus, one is free 
to apply a potentially more complete and reliable approach than scale variation to evaluating the theoretical uncertainty due to missing higher orders in the PDF fit. Indeed, our result may be taken as further evidence that such an approach is preferable. In such a case, the analysis above will not directly apply, although some element of the basic approach, namely expression of the predicted observables directly in terms of the fit observables, will certainly be relevant. If, on the other hand, one does apply the standard factorization scale variation approach, then clearly considerable care is necessary to maintain consistency with the requirements demonstrated in this paper. Future work will consider the impact of such variations, consistently performed, within the context of the global MMHT fit and its interplay with the tolerance criteria to evaluate the PDF uncertainties.

\section*{Acknowledgements}

LHL thanks the Science and Technology Facilities Council (STFC) for support via grant award ST/P004547/1.  RST thanks the Science and Technology Facilities Council (STFC) for support via grant award ST/P000274/1.
We would like to thank James Stirling for many illuminating discussions on this topic in particular, and many 
others besides over the years.

\appendix

\section{DGLAP evolution in the diagonal basis}\label{sec:ap1}

In this appendix we summarise a few standard formulae for the DGLAP evolution in a diagonal basis. The coupled DGLAP equation for the quark singlet $\Sigma_q$ and gluon $g$ PDF reads
\be\label{eq:dglapmat}
\frac{\partial}{\partial \ln \mu^2} \begin{pmatrix} g\\ \Sigma_q\end{pmatrix}=\tilde{\alpha}_S\begin{pmatrix}\gamma_{gg}&\gamma_{gq}\\ 2 n_f\gamma_{qg} & \gamma_{qq}\end{pmatrix}  \begin{pmatrix} g\\ \Sigma_q\end{pmatrix}\;,
\ee
in Mellin space, where we leave the scale and moment arguments implicit. This can be diagonalized to give
\be
\frac{\partial}{\partial \ln \mu^2} \begin{pmatrix} \Sigma_+\\ \Sigma_- \end{pmatrix}=\tilde{\alpha}_S\begin{pmatrix}\gamma_{+}&0\\ 0& \gamma_{-}\end{pmatrix}  \begin{pmatrix} \Sigma_+\\ \Sigma_-\end{pmatrix}\;,
\ee
allowing us to write
\be\label{eq:sigev}
\Sigma_{\pm}(j,\mu^2)=\Sigma_{\pm}(j,Q^2)\left(\frac{\mu^2}{Q^2}\right)^{\tilde{\alpha}_S\gamma_{\pm}}\;,
\ee
where
\begin{align}
\Sigma_+(j,\mu^2) &= \frac{1}{r_- - r_+}\left[ r_- g(j,\mu^2) - \Sigma_q(j,\mu^2)\right]\;,\\
\Sigma_-(j,\mu^2) &=  \frac{1}{r_- - r_+}\left[ -r_+ g(j,\mu^2) + \Sigma_q(j,\mu^2)\right]\;,
\end{align}
with
\be
\gamma_\pm = \frac{1}{2}\left[ \gamma_{gg} + \gamma_{qq} \pm \sqrt{(\gamma_{gg} - \gamma_{qq})^2 + 8 n_f \gamma_{qg}\gamma_{gq}}\right] \;, \qquad r_\pm = \frac{2n_f\gamma_{qg}}{\gamma_\pm - \gamma_{qq}}\;.
\ee
The LO splitting functions are given by
\begin{align}
\gamma_{qq}(j) &= C_F \left[-\frac{1}{2} + \frac{1}{j(j+1)} - 2\sum_{k=2}^j \frac{1}{k}\right]\;,\\
\gamma_{qg}(j) &= T_R \left[\frac{(2+j+j^2)}{j(j+1)(j+2)}\right]\;,\\
\gamma_{gq}(j) &= C_F \left[\frac{(2+j+j^2)}{j(j+1)(j-1)}\right]\;,\\
\gamma_{gg}(j) &= 2 N_C \left[-\frac{1}{12} + \frac{1}{j(j-1)} + \frac{1}{(j+1)(j+2)} - \sum_{k=2}^j \frac{1}{k}\right]-\frac{2}{3}n_F T_R\;,
\end{align}
while the NLO case is given in~\cite{GonzalezArroyo:1979he}.
In the high $x$ ($j\gg 1$) limit, we then have at LO
\be
\gamma_{qq} = -2 C_F \ln j \;, \quad \gamma_{gg} = -2N_C \ln j\;,\quad \gamma_{qg},\gamma_{gq} = 0\;,
\ee
with the same scaling present at NLO. In other words, \eqref{eq:dglapmat} is already diagonal and we can simply write
\be
\Sigma_+(j,\mu^2)  = g(j,\mu^2) \;, \qquad \Sigma_-(j,\mu^2) = \Sigma_{q}(j,\mu^2)\;,
\ee
up to overall normalisation factors.
 In the low $x$ ($j \sim 1$) limit (see~\cite{Blumlein:1997em} for further discussion), we have at LO
\be
\gamma_{gq}= \frac{2 C_F}{j-1}\;,\quad \gamma_{gg} = \frac{2 N_C}{j-1}\;, \quad \gamma_{qq} = 0\;,\quad \gamma_{qg} =\frac{2}{3}T_R \;,
\ee
so that 
\be
\gamma_+ = \gamma_{gg}\sim \frac{1}{j-1}\;,\qquad \gamma_- = -2n_f\frac{\gamma_{qg}\gamma_{gq}}{\gamma_{gg}}\sim {\rm const.}\;,
\ee
and
\be
r_+ = \frac{1}{9}n_f (j-1)\;,\qquad r_- = -\frac{9}{4}\;.
\ee
Thus, if we consider the evolution of the quark singlet and gluon at the low input scale $\mu_0\sim $ 1 GeV, and a larger scale $\mu$, e.g. characteristic of an LHC process, we have
\be 
g(j,\mu^2) = \Sigma_+(j,\mu^2_0)\left(\frac{\mu^2}{\mu_0^2}\right)^{\tilde{\alpha}_S \gamma_{+}} +\Sigma_-(j,\mu^2_0)\overset{\mu^2\gg \mu_0^2}{\to} \Sigma_+(j,\mu^2_0)\left(\frac{\mu^2}{\mu_0^2}\right)^{\tilde{\alpha}_S \gamma_{+}} \:,
\ee
and
\begin{align}\nonumber
\Sigma_q(j,\mu^2) &=  -\frac{9}{4}\Sigma_-(j,\mu^2_0) + \frac{1}{9}n_f(j-1) \Sigma_+(j,\mu^2_0)\left(\frac{\mu^2}{\mu_0^2}\right)^{\tilde{\alpha}_S \gamma_{+}}\;,\\ \nonumber
&= \Sigma_q(j,\mu_0^2) + \frac{1}{9}n_f(j-1) \biggl( g(j,\mu_0^2) +\frac{4}{9}\Sigma_q(j,\mu_0^2)
\biggr)\biggl[\left(\frac{\mu^2}{\mu_0^2}\right)^{\tilde{\alpha}_S \gamma_{+}}-1
\biggr] \;,\\
&\overset{\mu^2\gg \mu_0^2}{\to} \frac{1}{9}n_f(j-1) \biggl( g(j,\mu_0^2) +\frac{4}{9}\Sigma_q(j,\mu_0^2)
\biggr)\left(\frac{\mu^2}{\mu_0^2}\right)^{\tilde{\alpha}_S \gamma_{+}}\;.
\end{align}
Thus at high scale the positive eigenvector dominates for both the gluon and quark singlet, in the latter case overcoming the normalization due to the power of $j-1$ compared to the negative eigenvector input. At NLO, the $\gamma_-$ eigenvector picks up a $\sim \alpha_S/(j-1)$ pole, but this is nonetheless suppressed by a power of $\alpha_S$ relative to $\gamma_+$ and thus the leading high $\mu^2$ scaling is still driven by the latter eigenvector, with the $O(\alpha_S)$ coefficient of the $\Sigma_+$ contribution $\Sigma_q$ being non--vanishing as $j \to 1$. 

We therefore have that for sufficiently high scales an observable will be given dominantly by the $\Sigma_+$ eigenvector, with $g \sim \Sigma_q$ due to the low--$x$ PDF evolution. Alternatively, if the our observable is given by the rate of change with scale $Q^2$, e.g. ${\rm d}F_2/{\rm d}Q^2$, then the input $\Sigma_-$ component will give no contribution, and the positive eigenvector will be dominant at all scales.

\appendix

\bibliography{references}{}

\begin{thebibliography}{10}

\bibitem{Gao:2017yyd}
J.~Gao, L.~Harland-Lang, and J.~Rojo,
\newblock Phys. Rept. {\bf 742}, 1 (2018), 1709.04922.

\bibitem{Abramowicz:2015mha}
ZEUS, H1, H.~Abramowicz {\em et~al.},
\newblock Eur. Phys. J. {\bf C75}, 580 (2015), 1506.06042.

\bibitem{Abe:1996wy}
CDF, F.~Abe {\em et~al.},
\newblock Phys. Rev. Lett. {\bf 77}, 438 (1996), hep-ex/9601008.

\bibitem{Huston:1995tw}
J.~Huston {\em et~al.},
\newblock Phys. Rev. Lett. {\bf 77}, 444 (1996), hep-ph/9511386.

\bibitem{Pumplin:2001ct}
J.~Pumplin {\em et~al.},
\newblock Phys. Rev. {\bf D65}, 014013 (2001), hep-ph/0101032.

\bibitem{Pumplin:2002vw}
J.~Pumplin {\em et~al.},
\newblock JHEP {\bf 0207}, 012 (2002), hep-ph/0201195.

\bibitem{Martin:2002aw}
A.~D. Martin, R.~G. Roberts, W.~J. Stirling, and R.~S. Thorne,
\newblock Eur. Phys. J. {\bf C28}, 455 (2003), hep-ph/0211080.

\bibitem{Alekhin:1996za}
S.~Alekhin,
\newblock Eur. Phys. J. {\bf C10}, 395 (1999), hep-ph/9611213.

\bibitem{Botje:1999dj}
M.~Botje,
\newblock Eur. Phys. J. {\bf C14}, 285 (2000), hep-ph/9912439.

\bibitem{Barone:1999yv}
V.~Barone, C.~Pascaud, and F.~Zomer,
\newblock Eur. Phys. J. {\bf C12}, 243 (2000), hep-ph/9907512.

\bibitem{Giele:2001mr}
W.~T. Giele, S.~A. Keller, and D.~A. Kosower,
\newblock (2001), hep-ph/0104052.

\bibitem{mvvns}
S.~Moch, J.~A.~M. Vermaseren, and A.~Vogt,
\newblock Nucl. Phys. {\bf B688}, 101 (2004), hep-ph/0403192.

\bibitem{Vogt:2004mw}
A.~Vogt, S.~Moch, and J.~A.~M. Vermaseren,
\newblock Nucl. Phys. {\bf B691}, 129 (2004), hep-ph/0404111.

\bibitem{Martin:2007bv}
A.~D. Martin, W.~J. Stirling, R.~S. Thorne, and G.~Watt,
\newblock Phys. Lett. {\bf B652}, 292 (2007), 0706.0459.

\bibitem{Alekhin:2009ni}
S.~Alekhin, J.~Blumlein, S.~Klein, and S.~Moch,
\newblock Phys. Rev. {\bf D81}, 014032 (2010), 0908.2766.

\bibitem{Ball:2017nwa}
NNPDF, R.~D. Ball {\em et~al.},
\newblock Eur. Phys. J. {\bf C77}, 663 (2017), 1706.00428.

\bibitem{Butterworth:2015oua}
J.~Butterworth {\em et~al.},
\newblock J. Phys. {\bf G43}, 023001 (2016), 1510.03865.

\bibitem{Watt:2012tq}
G.~Watt and R.~S. Thorne,
\newblock JHEP {\bf 08}, 052 (2012), 1205.4024.

\bibitem{Carrazza:2015aoa}
S.~Carrazza, S.~Forte, Z.~Kassabov, J.~I. Latorre, and J.~Rojo,
\newblock Eur. Phys. J. {\bf C75}, 369 (2015), 1505.06736.

\bibitem{Andersen:2014efa}
J.~R. Andersen {\em et~al.},
\newblock (2014), 1405.1067.

\bibitem{Thorne:2012az}
R.~S. Thorne,
\newblock Phys. Rev. {\bf D86}, 074017 (2012), 1201.6180.

\bibitem{Ball:2013gsa}
NNPDF, R.~D. Ball {\em et~al.},
\newblock Phys. Lett. {\bf B723}, 330 (2013), 1303.1189.

\bibitem{Thorne:2014toa}
R.~S. Thorne,
\newblock Eur. Phys. J. {\bf C74}, 2958 (2014), 1402.3536.

\bibitem{Martin:2009iq}
A.~D. Martin, W.~J. Stirling, R.~S. Thorne, and G.~Watt,
\newblock Eur.Phys.J. {\bf C63}, 189 (2009), 0901.0002.

\bibitem{Pearson:2018tim}
R.~L. Pearson and C.~Voisey,
\newblock {Towards parton distribution functions with theoretical
  uncertainties},
\newblock 2018, 1810.01996.

\bibitem{Altarelli:1979ub}
G.~Altarelli, R.~K. Ellis, and G.~Martinelli,
\newblock Nucl. Phys. {\bf B157}, 461 (1979).

\bibitem{Grunberg:1982fw}
G.~Grunberg,
\newblock Phys. Rev. {\bf D29}, 2315 (1984).

\bibitem{Catani:1996sc}
S.~Catani,
\newblock Z. Phys. {\bf C75}, 665 (1997), hep-ph/9609263.

\bibitem{Thorne:1997mb}
R.~S. Thorne,
\newblock Nucl. Phys. {\bf B512}, 323 (1998), hep-ph/9710541.

\bibitem{Blumlein:2000wh}
J.~Blumlein, V.~Ravindran, and W.~L. van Neerven,
\newblock Nucl. Phys. {\bf B586}, 349 (2000), hep-ph/0004172.

\bibitem{Hentschinski:2013zaa}
M.~Hentschinski and M.~Stratmann,
\newblock (2013), 1311.2825.

\bibitem{Davies:2017hyl}
J.~Davies and A.~Vogt,
\newblock Phys. Lett. {\bf B776}, 189 (2018), 1711.05267.

\bibitem{SF}
Stefano Forte,
  \texttt{http://nnpdf.mi.infn.it/wp-content/uploads/2018/07/SF\_cernth18.pdf}.

\bibitem{Currie:2018xkj}
J.~Currie {\em et~al.},
\newblock Submitted to: JHEP  (2018), 1807.03692.

\bibitem{Lindert:2017olm}
J.~M. Lindert {\em et~al.},
\newblock Eur. Phys. J. {\bf C77}, 829 (2017), 1705.04664.

\bibitem{Cacciari:2011ze}
M.~Cacciari and N.~Houdeau,
\newblock JHEP {\bf 09}, 039 (2011), 1105.5152.

\bibitem{David:2013gaa}
A.~David and G.~Passarino,
\newblock Phys. Lett. {\bf B726}, 266 (2013), 1307.1843.

\bibitem{Bagnaschi:2014wea}
E.~Bagnaschi, M.~Cacciari, A.~Guffanti, and L.~Jenniches,
\newblock JHEP {\bf 02}, 133 (2015), 1409.5036.

\bibitem{Gao:2013xoa}
J.~Gao {\em et~al.},
\newblock Phys. Rev. {\bf D89}, 033009 (2014), 1302.6246.

\bibitem{GonzalezArroyo:1979he}
A.~Gonzalez-Arroyo and C.~Lopez,
\newblock Nucl. Phys. {\bf B166}, 429 (1980).

\bibitem{Blumlein:1997em}
J.~Blumlein and A.~Vogt,
\newblock Phys. Rev. {\bf D58}, 014020 (1998), hep-ph/9712546.

\end{thebibliography}
\bibliographystyle{h-physrev}

\end{document}